\documentclass[epjCONF]{svjour}
\usepackage{graphics,amsmath,amssymb,xspace,textcomp,setspace,multirow,lineno}
\usepackage[rflt]{floatflt}
\usepackage[varg]{txfonts} 
\usepackage[latin1]{inputenc}
\newcommand{\xmax}{{$X_\text{max}$}\xspace}

\def \gcmsq    {$\text{g~cm}^{-2}$\xspace}

\def \ntwo {N$_2$\xspace}
\def \ntwoplus {N$_2^+$\xspace}

\def \pprimevapor {$p^\prime_{\rm{H}_2\rm{O}}$\xspace}
\def \pprimeair {$p^\prime_{\rm{air}}$\xspace}
\def \setRefyield {S$_{\rm{FY}_{\rm{Ref}}}$\xspace}
\def \setAugNag {S$_{\rm{Auger\_NaganoScale}}$\xspace}
\def \setTANag {S$_{\rm{TA\_NaganoScale}}$\xspace}
\def \setTAFlash {S$_{\rm{TA-FLASHScale}}$\xspace}
\def \setRefyieldAirfly {S$_{\rm{FY}_{\rm{Ref}}-\rm{AIRFLYScale}}$\xspace}
\def \setRefyieldTUM {S$_{\rm{FY}_{\rm{Ref}}-\rm{TUMScale}}$\xspace}
\def\Offline{\mbox{$\overline{\textrm%
{Off}}$\hspace{.05em}\protect\raisebox{.4ex}%
{$\protect\underline{\textrm{line}}$}}\xspace}

\session-title{UHECR2012 Symposium}
\begin{document}
\title{Nitrogen fluorescence in air for observing extensive air showers}
\author{B.~Keilhauer\inst{1}\fnmsep\thanks{\email{bianca.keilhauer@kit.edu}} 
       \and M.~Bohacova\inst{2} 
       \and M.~Fraga\inst{3} 
       \and J.~Matthews\inst{4} 
       \and N.~Sakaki~\inst{1,5} 
       \and Y.~Tameda\inst{6} 
       \and Y.~Tsunesada\inst{7} 
       \and A.~Ulrich\inst{8}}
\institute{Karlsruhe Institute of Technology (KIT), Karlsruhe, Germany 
\and Institute of Physics (FZU) of the Academy of Sciences, 
     Prague, Czech Republic 
\and LIP-Coimbra and Departamento de F\'isica, Universidade de Coimbra, Portugal 
\and University of Utah, USA 
\and Aoyama Gakuin University, Japan 
\and University of Tokyo, Japan 
\and Tokyo Institute of Technology, Japan 
\and Technische Universit\"at M\"unchen, Germany}
\abstract{
Extensive air showers initiate the fluorescence emissions from nitrogen 
mole\-cules in air. The UV-light is emitted isotropically and can be used 
for observing the longitudinal development of extensive air showers in the 
atmosphere over tenth of kilometers. This measurement technique is well-established 
since it is exploited for many decades by several cosmic ray experiments. 
However, a fundamental aspect of the air shower analyses is the description 
of the fluorescence emission in dependence on varying atmospheric conditions. 
Different fluorescence yields affect directly the energy scaling of air shower 
reconstruction. In order to explore the various details of the nitrogen 
fluorescence emission in air, a few experimental groups have been performing 
dedicated measurements over the last decade. Most of the measurements 
are now finished. These experimental groups have been discussing their 
techniques and results in a series of \emph{Air Fluorescence Workshops} 
commenced in 2002. \newline
At the 8$^{\rm{th}}$ Air Fluorescence Workshop 2011, it was suggested to 
develop a common way of describing the nitrogen fluorescence for application 
to air shower observations. Here, first analyses for a common treatment of 
the major dependences of the emission procedure are presented. Aspects like 
the contributions at different wavelengths, the dependence on pressure as 
it is decreasing with increasing altitude in the atmosphere, the temperature 
dependence, in particular that of the collisional cross sections between 
molecules involved, and the collisional de-excitation by water vapor are discussed.
} 
\maketitle
%

\section{Introduction}
\label{sec:intro}

The detection of ultra-high energy cosmic rays is very challenging and it is
a major key to answer open questions in astroparticle physics. Cosmic rays
are messengers of processes in their sources and at their propagation towards
us. Two important aspects of the cosmic ray properties are their energy and
mass composition in order to evaluate the shape of the cosmic ray spectrum
\cite{Berezinsky2012}.

For detecting ultra-high energy cosmic rays, extensive air showers initiated
in the Earth's atmosphere are observed. Manifold techniques have been 
developed over the last decades and in this article, we will focus on the 
fluorescence technique.

While extensive air showers are developing through the
atmosphere, their energy deposit causes an excitation of atmospheric 
nitrogen molecules. The spontaneous de-excitation can be separated into 
a radiative and a non-radiative channel. The isotropically emitted 
UV fluorescence light is proportional to the energy deposit 
and can be used to perform an almost calorimetric measurement of the
air shower energy \cite{5AFW_summary}. The fluorescence technique was
investigated initially in \cite{bunner1967} and found its successful
application at Fly's Eye \cite{baltrusaitis1985}, HiRes
(High Resolution Fly's Eye experiment) \cite{hires2000},
the Pierre Auger Observatory \cite{augerFD2010}, and TA (Telescope
Array) \cite{TA2008}. With JEM-EUSO
(Extreme Universe Space Observatory onboard the Japanese Experiment Module), 
there are even plans to 
put this fluorescence technique onboard the ISS\footnote{International 
Space Station} to observe extremely high-energy cosmic rays from
space~\cite{jem-euso}.

For determining the absolute fluorescence yield, its emission spectrum, as well
as its dependence on varying atmospheric conditions to an appropriate
level of precision, several experimental groups have been performing 
dedicated laboratory measurements over the last decade. The experimental
efforts have been supplemented by several simulations and air
shower reconstruction studies. These groups have been discussing their 
techniques and data in a series of \emph{Air Fluorescence Workshops} (AFW)
commenced in Utah, 2002 \cite{8afw2011}. Main results are contributed
by the following alphabetic list of collaborations and groups led by
their principal investigators: AIRFLY 
\cite{airfly1,airfly2,airfly3,airfly4,airfly5,airfly6}, 
Airlight \cite{airlight1,airlight2}, 
Arqueros/UCM \cite{UCM1,UCM2,UCM3,UCM4,UCM5}, 
FLASH \cite{flash1,flash2,flash3,flash4}, Fraga/LIP \cite{LIP1,LIP2}, 
Keilhauer/KIT \cite{KIT1,KIT2,KIT3}, Lefeuvre/APC \cite{APC}, 
MACFLY \cite{macfly}, Nagano/Sakaki \cite{Nagano2003,Nagano2004,Sakaki2008}, 
and Ulrich/TUM \cite{TUM1,TUM2,TUM3,TUM4}.
At the up to now last \emph{Air Fluorescence Workshop} in Karlsruhe, 2011,
it was decided to develop a common fluorescence description for application
at air shower reconstructions \cite{8afw2011}. The description of the 
fluorescence emission
can be summarized by three different aspects -- the absolute fluorescence
yield, e.g. at a representative band head, the wavelength-dependent 
spectrum, and the dependence on atmospheric conditions. At a first step, the two
dependences are being discussed to be able to compare reconstructions of
extensive air showers based on a common altitude-dependent shape of
the fluorescence emission. In a second step, the challenge of
determining the absolute fluorescence yield is being addressed
which directly introduces the scaling of the reconstructed primary
energy of air showers.

\section{Description of the fluorescence emission from molecular nitrogen in air}
\label{sec:fl_descr}

The fluorescence light initiated by extensive air showers is mainly produced
by the energy deposit of electrons/positrons from the air shower in 
inelastic collisions with air molecules. The observed emission is dominated
by radiative de-excitation of \ntwo and \ntwoplus in the wavelength
range between about 290 and 430~nm. The radiative de-excitation is competing
with quenching processes caused by further non-excited molecules in air.
Thus, the fluorescence yield $Y_{\rm air}$, given as the number of photons 
emitted per 
unit deposited energy, can be determined by the intrinsic fluorescence
yield $Y_{\rm air}^0$ at zero pressure reduced by the non-radiative de-excitation.
Here, we follow the nomenclature as defined in \cite{5AFW_summary}. The 
de-excitation processes can either be described by the natural lifetime 
of an excited state together with the quenching rate constant or 
collisional cross sections (cf.\ \cite{bunner1967}, 
\cite{airlight1,airlight2}, \cite{TUM1,TUM2,TUM3,TUM4}) or by introducing 
a reference pressure $p^\prime$,
at which the probability of collisional quenching equals that of radiative 
de-excitation, in the Stern-Volmer factor \cite{stern1919}.

%
In this monograph, we employ the formulation using the Stern-Volmer factor 
for the fluorescence yield $Y_{\rm air}$ 
as a function of the photon wavelength $\lambda$ 
and atmospheric conditions $(p - {\rm{pressure}}, T - {\rm{temperature}}, 
e - {\rm{water~vapor~pressure}})$ 
\begin{eqnarray}
Y_{\rm air}(\lambda, p, T) &=& Y_{\rm air}(337{\rm nm}, p_0, T_0) \cdot 
I_{\lambda}/I_{\rm 337}(p_0, T_0) \cdot 
\frac{1 + \frac{p_0}{p'_{\rm air}(\lambda, T_0)}}{1 + \frac{p}{p'_{\rm air}(\lambda, T_0) 
\cdot \sqrt{\frac{T}{T_0}}  \cdot \frac{H_{\rm \lambda}(T_0)}{H_{\lambda}(T)}}} 
\label{eq:flyield}
\end{eqnarray}
and taking into account the temperature-dependent collisional cross
sections between excited nitrogen and air molecules in general by
\begin{eqnarray}
\frac{H_{\lambda}(T)}{H_{\lambda}(T_0)} = 
\left( \frac{T}{T_0}  \right)^{\alpha_{\lambda}}. 
\label{eq:tempdep}
\end{eqnarray}
The reference atmospheric condition at which the 
laboratory fluorescence measurements have been carried out is denoted by $(p_0, T_0)$.
The wavelength dependence of the fluorescence intensity is expressed as 
$I_{\lambda}$ which is relative to that wavelength at which the absolute
yield is determined. The leading factor of the right 
side of Eq.~(\ref{eq:flyield}), $Y_{\rm air}(337{\rm nm})$, 
gives the absolute yield at $\lambda = 337.1$~nm where
the emission from \ntwo is strongest in the observation range. In 
Eq.~(\ref{eq:tempdep}), $\alpha_{\lambda}$
is the exponent of the power law describing the temperature-dependent 
collisional cross sections for each wavelength. 

For humid air, the characteristic pressure for dry air $p'_{\rm air}$ has 
to be substituted in the denominator of Eq.~(\ref{eq:flyield}) 
by
\begin{eqnarray}
 \frac{1}{p'_{\rm air}} \longrightarrow \frac{1}{p'_{\rm air}} 
\left(1 - \frac{p_h}{p} \right) + \frac{1}{p'_{\rm H_2 O}} \frac{e}{p}
\label{eq:humdep}
\end{eqnarray}
where $p'_{\rm H_2 O}$ 
is the characteristic pressure for collisional quenching with water vapor
and $p_h$ is the partial pressure of water vapor in the atmosphere.

Above, the different factors are written in dependence on $\lambda$. However,
the spectral emissions are mainly vibrational transitions from the 
Second Positive (2P) system of \ntwo and First Negative (1N) of \ntwoplus  
with some minor contributions from the Gaydon-Herman (GH) system
of \ntwo. Since the rotational structure of the molecular bands is 
not resolved in our experiments, the transitions can be described by
molecular bands\footnote{$v,~v^\prime$ as vibrational
quantum numbers of the upper and lower states} $v-v^\prime$ with a 
spectral width and shape determined by the rotational structure. Concerning
the spectral intensity of the emissions $I_{\lambda}$, each transition
between $v-v^\prime$ corresponds to one wavelength $\lambda$ and has,
of course, its individual strength. For the factors describing the 
quenching effects, however, only the physical properties of the upper, 
excited state are of relevance. Thus, the characteristic pressures
$p'_{\rm air}$ and $p'_{\rm H_2 O}$ as well as the exponent $\alpha$
are the same for all transitions arising from the same upper 
vibrational level~$v$.

\section{First draft of a common reference yield}
\label{sec:ref_yield}

As a first step, we describe the spectral and atmospheric dependences
of the fluorescence emissions, see Sec.~\ref{sec:intro}. This provides
us a common altitude-dependent shape of the fluorescence profile. For
application to air shower reconstructions, this requires an adequate
knowledge of atmospheric conditions at the place and time of an
observed air shower event \cite{KIT3}.

The values for all necessary parameters to describe the fluorescence
emission in air with the suggested procedure are summarized in 
Tab.~\ref{tab:refYield} and will be discussed in more details in the
following.

\begin{table}[htbp]
\caption{Input parameters for the common fluorescence description.}
\label{tab:refYield}
\renewcommand{\thefootnote}{\thempfootnote}
\renewcommand{\multirowsetup}{\centering}
\begin{minipage}{\linewidth}
\begin{center}
\begin{tabular}{ccccccc}\hline
system & band & $\lambda$ & $I_\lambda/I_{\rm 337}$ & $p'_{\rm air}$ & 
$p'_{\rm H_2 O}$ & $\alpha$ \\
 & & (nm) & (\%) & (hPa) & (hPa) & \\ \hline
\ntwo 2P & 0-0 & 337.1 & 100 & \multirow{4}{2cm}{15.83 $\pm$ 0.80} & \multirow{4}{2cm}{1.46 $\pm$ 0.05} & \multirow{4}{2cm}{-0.35 $\pm$ 0.08} \\
 & 0-1 & 357.7 & 67.4 $\pm$ 2.4 & & & \\
 & 0-2 & 380.5 & 27.2 $\pm$ 1.0 & & & \\
 & 0-3 & 405.0 & 8.07 $\pm$ 0.29 &  & & \\
\\
 & 1-0 & 315.9 & 39.3 $\pm$ 1.4 & \multirow{6}{2cm}{12.03 $\pm$ 0.66} & \multirow{6}{2cm}{1.90 $\pm$ 0.18} & \multirow{6}{2cm}{-0.20 $\pm$ 0.08} \\
 & 1-1 & 333.9 & 4.02 $\pm$ 0.18 & & & \\
 & 1-2 & 353.7 & 21.35 $\pm$ 0.76 & & & \\
 & 1-3 & 375.6 & 17.87 $\pm$ 0.63 & & & \\
 & 1-4 & 399.8 & 8.38  $\pm$ 0.29 & & & \\
 & 1-5 & 427.0 & 7.08  $\pm$ 0.28 & & & \\
\\
 & 2-0 & 297.7 & 2.77 $\pm$0.13 & \multirow{7}{2cm}{13.12 $\pm$ 0.71} & \multirow{7}{2cm}{1.80 $\pm$ 0.14} & \multirow{7}{2cm}{-0.17 $\pm$ 0.08} \\
 & 2-1 & 313.6 & 11.05 $\pm$ 0.41 & & & \\
 & 2-2 & 330.9 & 2.15 $\pm$ 0.12 & & & \\
 & 2-3 & 350.0 & 2.79 $\pm$ 0.11 & & & \\
 & 2-4 & 371.1 & 4.97 $\pm$ 0.22 & & & \\
 & 2-5 & 394.3 & 3.36 $\pm$ 0.15 & & & \\
 & 2-6 & 420.0 & 1.75 $\pm$ 0.10 & & & \\
\\
 & 3-1 & 296.2 & 5.16 $\pm$ 0.29 &  \multirow{5}{2cm}{19.88 $\pm$ 0.86} & \multirow{5}{2cm}{1.84 $\pm$ 0.2} & \multirow{5}{2cm}{-0.19 $\pm$ 0.08} \\
 & 3-2 & 311.7 & 7.24 $\pm$ 0.27 & & & \\
 & 3-3 & 328.5 & 3.80 $\pm$ 0.14 & & & \\
 & 3-5 & 367.2 & 0.54 $\pm$ 0.04 & & & \\
 & 3-7 & 414.1 & 0.49 $\pm$ 0.07 & & & \\
\\
 & 4-4 & 326.8 & 0.80 $\pm$ 0.08 &   \multirow{2}{2cm}{19 $\pm$ 5.0} & \multirow{2}{2cm}{1.84 $\pm$ 0.2} & \multirow{2}{2cm}{-0.19 $\pm$ 0.08} \\
 & 4-7 & 385.8 & 0.50 $\pm$ 0.08 & & & \\
\\
\ntwoplus 1N & 0-0 & 391.4 & 28.0 $\pm$ 1.0 &  \multirow{2}{2cm}{2.94 $\pm$ 0.33} & \multirow{2}{2cm}{0.47 $\pm$ 0.02} & \multirow{2}{2cm}{-0.76 $\pm$ 0.08} \\
 & 0-1 & 427.8 & 4.94 $\pm$ 0.19 & & & \\
\\
 & 1-1 & 388.5 & 0.83 $\pm$ 0.04 & \multirow{2}{2cm}{3.92 $\pm$ 0.32} & \multirow{2}{2cm}{0} & \multirow{2}{2cm}{0} \\
 & 1-2 & 423.6 & 1.04 $\pm$ 0.11 & & & \\
\\
\ntwo GH & 0-4 & 346.3 & 1.74 $\pm$ 0.11 & \multirow{3}{2cm}{7.98 $\pm$ 0.56} & \multirow{3}{2cm}{0} & \multirow{3}{2cm}{0} \\
 & 0-5 & 366.1 & 1.13 $\pm$ 0.08 & & & \\
 & 0-6 & 387.7 & 1.17 $\pm$ 0.06 & & & \\
\\
 & 5-2 & 308.0 & 1.44 $\pm$ 0.10 & 21 $\pm$ 10.0 & 0 & 0 \\
\\
 & 6-2 & 302.0 & 0.41 $\pm$ 0.06 & \multirow{2}{2cm}{21 $\pm$ 10.0} & \multirow{2}{2cm}{0} & \multirow{2}{2cm}{0} \\
 & 6-3 & 317.6 & 0.46 $\pm$ 0.06 & & & \\ \hline
\end{tabular}
\end{center}
\end{minipage}
\end{table}

\subsection{Spectral intensities}
\label{sec:intensities}

\begin{figure}[htbp]
\hfill
\begin{minipage}[c]{0.48\textwidth}
\centering
\resizebox{1.\columnwidth}{!}{
\includegraphics{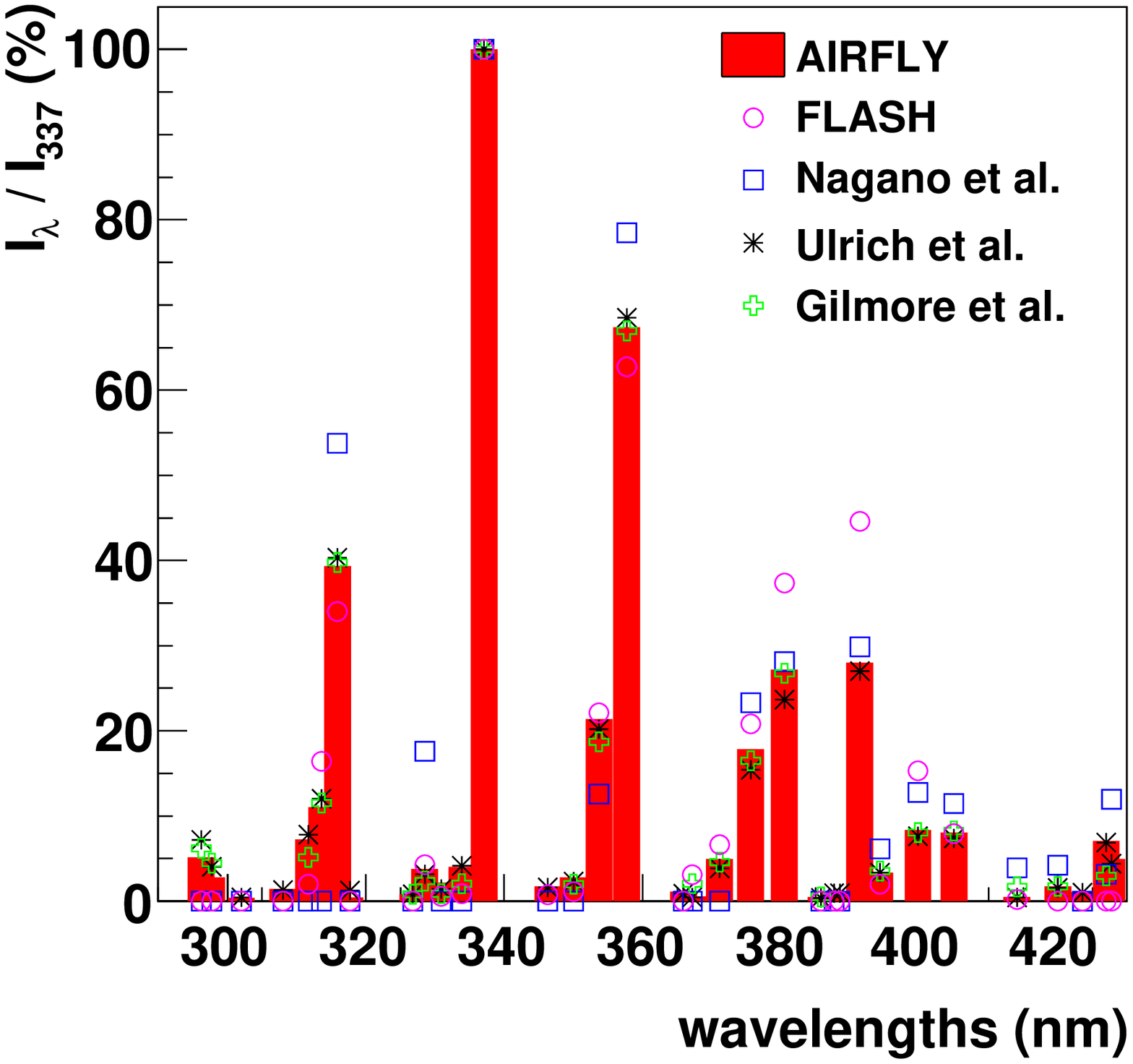}}
\end{minipage}
\hfill
\begin{minipage}[c]{0.48\textwidth}
\caption{Relative fluorescence intensities between about 300 and 430~nm. The sum
of the fluorescence yield in this wavelength range differs by -1.66\% 
(Ulrich et al.~\cite{TUM4}, by +2.08\% (Nagano et 
al.~\cite{Nagano2004}), and by -1.7\% 
(FLASH~\cite{flash2}) compared
to the sum of the fluorescence yield obtained by AIRFLY. The data from 
Gilmore et al.\ are calculated theoretically only for the emissions of the
2P system~\cite{gilmore}. Here, the sum of the fluorescence yield in the given wavelength 
range differs by -3.00\% compared to the yield of the 2P system obtained
by AIRFLY.}
\label{fig:intensities}
\end{minipage}
\end{figure}

For the values of $I_\lambda/I_{\rm 337}$, there is a quite complete
and precise measurement from AIRFLY \cite{airfly1,airfly2}. 
The spectral resolution of these data is 
very high which enables an adequate consideration of all dependences.
The data, which are adopted for the first suggestion of a common
reference yield, are displayed in Fig.~\ref{fig:intensities}, along 
with data from further working groups. A similar set, in terms of 
resolution and accuracy, of wavelength-
dependent intensities is obtained by \cite{TUM4}.
However for the sake of consistency, the data set measured by AIRFLY was 
chosen for this fluorescence description. 

\subsection{Atmospheric dependences}
\label{sec:atmos-dep}

The AIRFLY collaboration has published a complete set of \pprimeair($\lambda,T_0$)
values in dry air, one for each band~\cite{airfly1,airfly2}. As stated above, 
it is expected that all 
individual transitions starting at the same upper vibrational level have the 
same \pprimeair. Therefore, the suggestion is to use weighted
averages of the \pprimeair values measured for bands of a particular molecular
system with the same starting vibrational level. As weight factors, the quoted
uncertainties are used. Because of the experimental technique, AIRFLY was 
running, for all transitions apart from the 337.1~nm line, there is not only
a given uncertainty of the measurement itself, $\sigma_{p^\prime_{\rm{air}}}^{stat}$,
but an additional value for the propagated uncertainty of the measurement of the 
337.1~nm line, $\sigma_{p^\prime_{\rm{air}}}^{prop\_337}$. 
To take into account both uncertainties, a minimization of $\chi^2$ with
\begin{align}
\chi^2 = & \underbrace{\frac{(p^\prime_{\rm{air},00} - \mu_{0})^2}{\Bigl(\sigma_{p^\prime_{\rm{air},00}}^{stat}\Bigr)^2}}
+ \sum_{v^\prime=1}^{N_{lines}(0)} 
\frac{\biggl(p^\prime_{\rm{air},0v^\prime} - \mu_{0} + n \cdot \sigma_{p^\prime_{\rm{air},0v^\prime}}^{prop\_337}\biggr)^2}{\Bigl(\sigma_{p^\prime_{\rm{air},0v^\prime}}^{stat}\Bigr)^2}
+ \sum_{v=1}^{N_{bands}} \sum_{v^\prime=0}^{N_{lines}(v)} 
\frac{\biggl(p^\prime_{\rm{air},vv^\prime} - \mu_{v} + n \cdot \sigma_{p^\prime_{\rm{air},vv^\prime}}^{prop\_337}\biggr)^2}
     {\Bigl(\sigma_{p^\prime_{\rm{air},vv^\prime}}^{stat}\Bigr)^2} \notag \\
& ~~~~ := n^2 
\label{eq:chisquare}
\end{align}
is performed. The first term corresponds to the main transition \ntwo
2P(0,0) at 337.1~nm, the second term to all other transitions of the
2P(0,$v^\prime$) progression, and the double sum accounts 
for all other bands. The searched values \pprimeair for the different
bands are denoted as $\mu_v$ in Eq.~(\ref{eq:chisquare}) and 
listed in Tab.~\ref{tab:refYield} in the fifth column. The corresponding
uncertainties are derived by maximizing the influence of 
$\sigma_{p^\prime_{\rm{air}}}^{prop\_337}$ to the remaining data by
setting $n = \pm 1$ in the second and third term of Eq.~(\ref{eq:chisquare}). 
For the band progressions 2P(4,$v^\prime$), GH(5,$v^\prime$), and GH(6,$v^\prime$), 
we follow the recommendation of AIRFLY~\cite{airfly1}.

The quenching by water vapor molecules in the atmosphere can be taken
into account by applying Eq.~(\ref{eq:humdep}) to Eq.~(\ref{eq:flyield}).
The additional reference pressure \pprimevapor is determined experimentally
at a given temperature $T_0$. The group of Nagano/Sakaki has performed 
systematic studies about the effect
of quenching excited nitrogen molecules by water vapor (cf.\ overview talk by N.~Sakaki
at the 8$^{\rm{th}}$ AFW 2011~\cite{sakakiKA}). They measured 
$p^\prime_{\text{H}_2{\text{O}}}(\lambda,T_0)$
for several wavelengths in two different ways. First, they analyzed the measurements of
the photon yield and in a second step, they derived 
$p^\prime_{\text{H}_2{\text{O}}}(\lambda,T_0)$ from lifetime measurements. The 
results are recalled in Tab.~\ref{tab:p_H2O}. As also 
discussed by A.~Ulrich (cf.\ overview talk at the 8$^{\rm{th}}$ AFW 
2011~\cite{ulrichKA}), a 
systematic discrepancy was found between these two procedures. Even though
the details are not fully understood yet, the overall uncertainties 
propagating into air shower reconstructions are sufficiently small. 
\begin{table}[htb]
\begin{minipage}{0.25\linewidth}
\caption{Collisional quenching reference pressures for
water vapor quenching at $T_0 = $293~K as obtained by \cite{Sakaki2007,sakakiKA}.
\label{tab:p_H2O}}
\end{minipage}\hfill
\begin{minipage}{0.73\linewidth}
\begin{center}
\begin{tabular}{lccccc}\hline
 & & \multicolumn{2}{c}{from photon yield} & \multicolumn{2}{c}{from lifetime} \\ 
Band & $\lambda$ & $p^\prime_{\text{H}_2{\text{O}}}$ & $\sigma_{p^\prime_{\text{H}_2{\text{O}}}}$ & $p^\prime_{\text{H}_2{\text{O}}}$ & $\sigma_{p^\prime_{\text{H}_2{\text{O}}}}$ \\
 &  (nm) &  (hPa) &  (hPa) & (hPa) & (hPa) \\ \hline
2P(0,0) & 337.1 & 1.36 & 0.07 & 1.68 & 0.13 \\
2P(0,1) & 357.7 & 1.23 & 0.12 & 1.83 & 0.18 \\
2P(0,2) & 380.5 & 2.08 & 0.34 & 2.01 & 0.27 \vspace{8pt} \\
2P(1,0) & 315.9 & 2.23 & 0.54 & 2.38 & 0.35 \\
2P(1,4) & 399.8 & 1.30 & 0.29 & 2.26 & 0.41 \vspace{8pt} \\
2P(2,2) & 330.9 & 1.95 & 0.49 & 2.88 & 0.39 \\
2P(2,4) & 371.1 & 1.62 & 0.28 & 1.59 & 0.2 \vspace{8pt} \\
1N(0,0) & 391.4 & 0.40 & 0.04 & 0.42 & 0.03 \\ 
1N(0,1) & 427.8 & 0.53 & 0.07 & 0.89 & 0.07 \\ \hline
\end{tabular}
\end{center}
\end{minipage}
\end{table}
Here, we suggest to average the weighted averages of the two
independent data sets, again by applying the uncertainties as 
weights. The procedure and the propagation of the uncertainties is
performed by
\begin{equation}
p^\prime_{\rm{H}_2\rm{O},v} = \frac{\sum\limits_{i} p^\prime_{\rm{H}_2\rm{O},\lambda_i} 
    \cdot \frac{1}{\bigl(\sigma_{p^\prime_{\rm{H}_2\rm{O},\lambda_i}}\bigr)^2}}
       {\sum\limits_{i}\frac{1 }{\bigl(\sigma_{p^\prime_{\rm{H}_2\rm{O},\lambda_i}}\bigr)^2}}
\label{eq:weighting}
\end{equation}
and
\begin{equation}
\sigma_{p^\prime_{\rm{H}_2\rm{O},v}} = \sqrt{\frac{1 }{\sum\limits_{i}\frac{1}{\bigl(\sigma_{p^\prime_{\rm{H}_2\rm{O},\lambda_i}}\bigr)^2}}},
\label{eq:error_weighting}
\end{equation}
respectively. The final values are listed in Tab.~\ref{tab:refYield}.
As indicated in Tab.~\ref{tab:p_H2O}, no data are given for several
weaker transitions. In this case, we suggest to apply a weighted average
of the \pprimevapor values of $v = 1$ and 2 for those of 
$v = 3$ and 4. The emissions for the two weaker bands 
2P(3,$v^\prime$) and 2P(4,$v^\prime$) contribute with 4.8\% to the entire
emission at atmospheric ground conditions (c.f.\ Tab.~\ref{tab:refYield}). 
For all other band systems, 
the \pprimevapor is set to zero because of missing information but these
transitions contribute with only 2.1\% to the total intensity (c.f.\ 
Tab.~\ref{tab:refYield}).

A set of measurements of the $\alpha$-coefficient describing the temperature
dependence of the quenching cross sections of the main bands of the 2P and
1N band systems in air is also provided by the AIRFLY
Collaboration~\cite{airfly4}. The most recent data have
been presented at the 6$^{\rm th}$ Air Fluorescence Workshop in L'Aquila, 
Italy~\cite{laquila}. Because of the same experimental technique, the 
$\alpha$-values are derived by following the same procedure as for \pprimeair.
In the absence of further experimental data for weaker transitions,
the weighted averages of $\alpha$ values of $v = 3$ and 4 are
calculated correspondingly to the procedure for \pprimevapor.

\begin{figure}[htb]
\hfill
\begin{minipage}[c]{0.48\textwidth}
\centering
\resizebox{1.\columnwidth}{!}{
\includegraphics{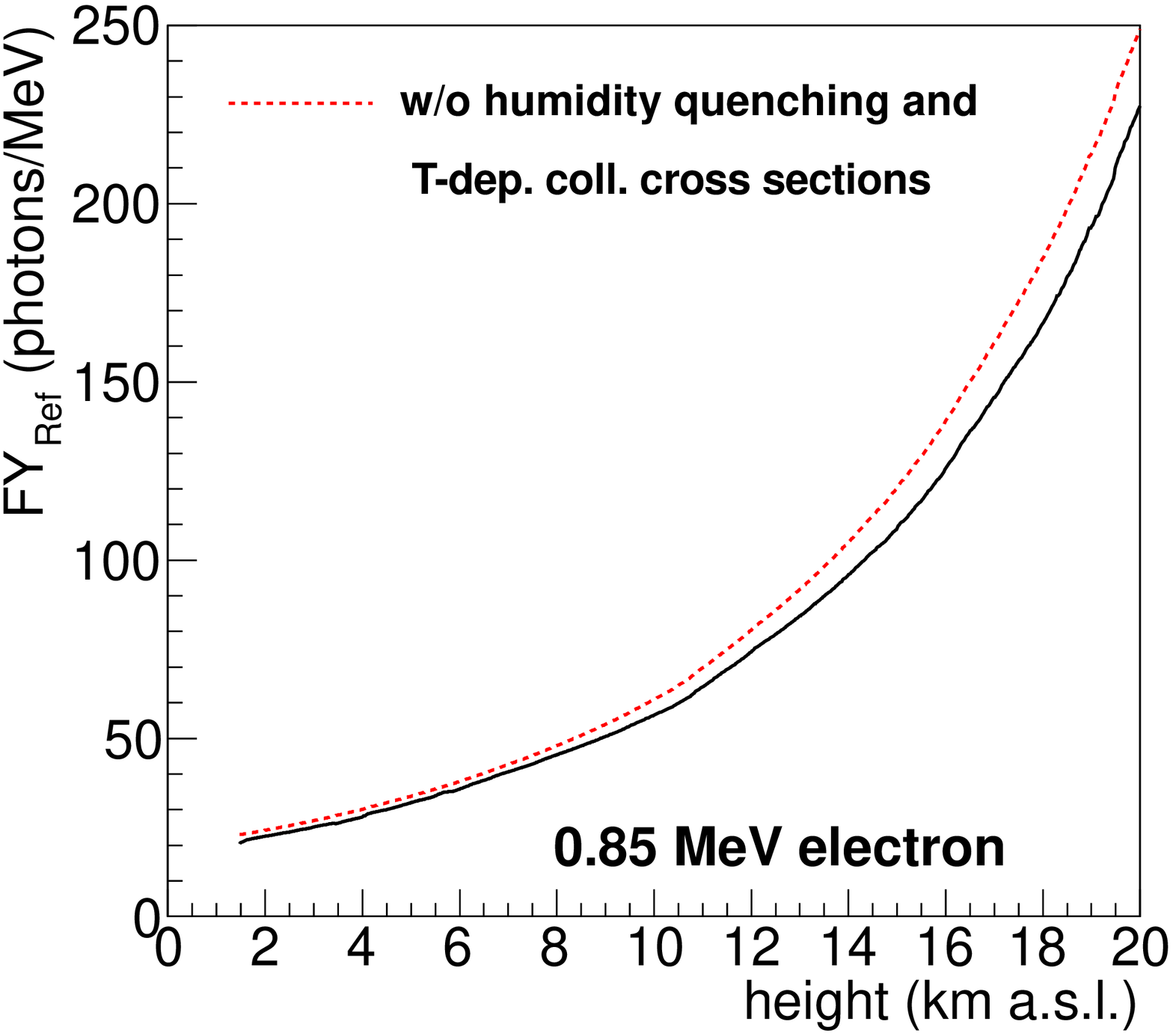}}
\end{minipage}
\hfill
\begin{minipage}[c]{0.48\textwidth}
\centering
\resizebox{1.\columnwidth}{!}{
\includegraphics{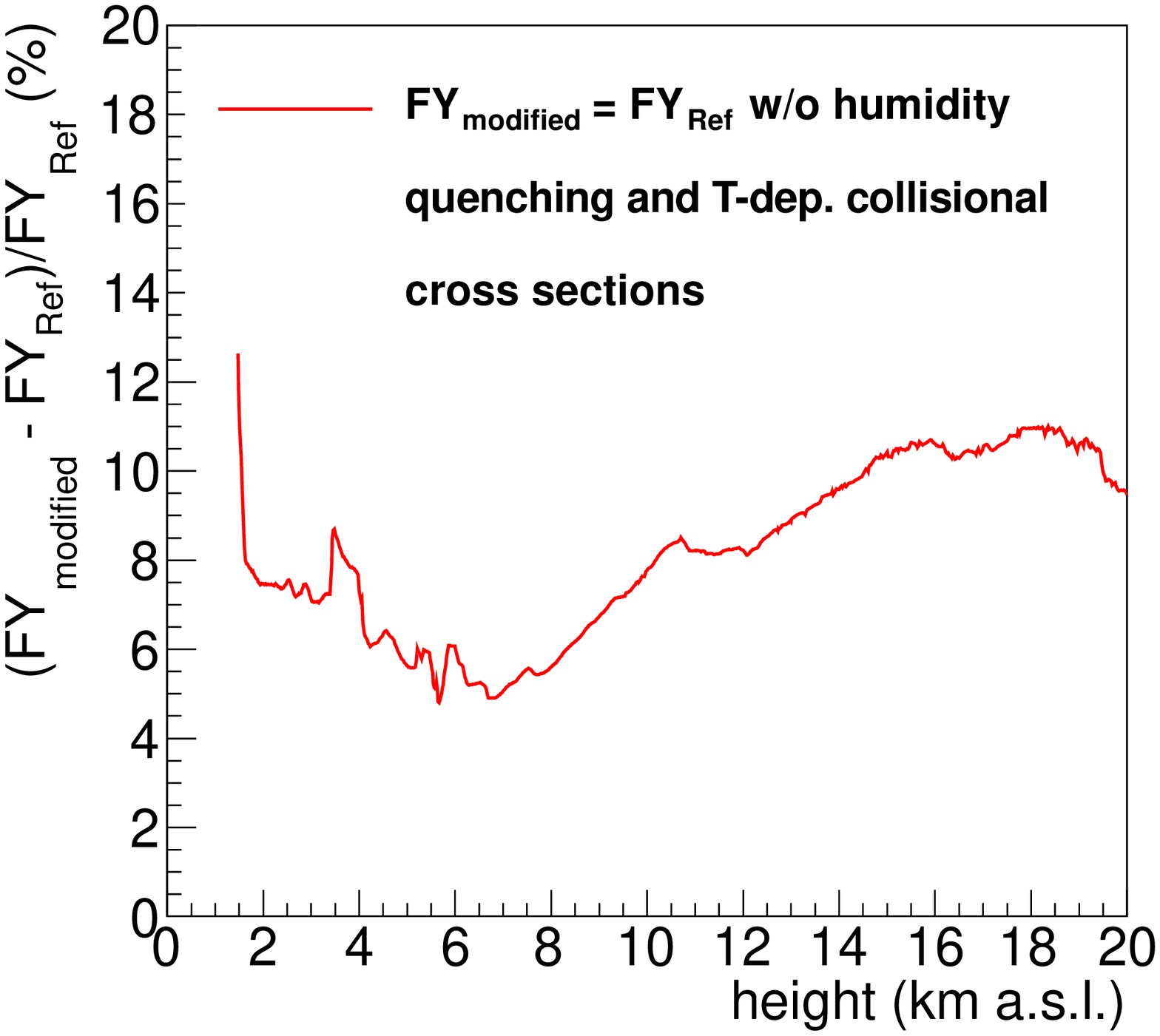}}
\end{minipage}
\caption{Left: Fluorescence yield for an ``always fresh'' 0.85~MeV electron. 
The absolute scale is determined by the fluorescence yield of the 
337.1~nm line at 800~hPa and 293~K taken from \cite{Nagano2004} which is
6.38~$\gamma_{337}$/MeV. The black, solid line represents the fluorescence
emission including all known atmospheric dependences. An exemplary 
atmospheric profile for early morning, 
summer conditions at the site of the Auger Observatory has been used. The ground
temperature was about 290~K and the relative humidity about 65\%.The red, dashed line
represents the emission as described by Eq.~\ref{eq:flyield} without taking 
into account Eq.~\ref{eq:humdep} and setting $\alpha_\lambda$ in Eq.~\ref{eq:tempdep}
to zero. Right: Difference of the modified to the standard reference description of
the fluorescence emission. The abrupt variations of the fluorescence yield are
due to small layers of increased water vapor in the given atmospheric profile.}
\label{fig:academic_FY}
\end{figure}

Having derived all necessary information to determine the fluorescence
emission in air, an academic fluorescence yield profile is plotted in
Fig.~\ref{fig:academic_FY} for an 0.85~MeV electron. In addition,
the fluorescence emission is plotted for the case of no water vapor
quenching nor applying the temperature-dependent collisional cross
sections (red, dashed line).

\begin{figure}[h!]
\hfill
\begin{minipage}[t]{0.48\textwidth}
\centering
\resizebox{1.\columnwidth}{!}{
\includegraphics{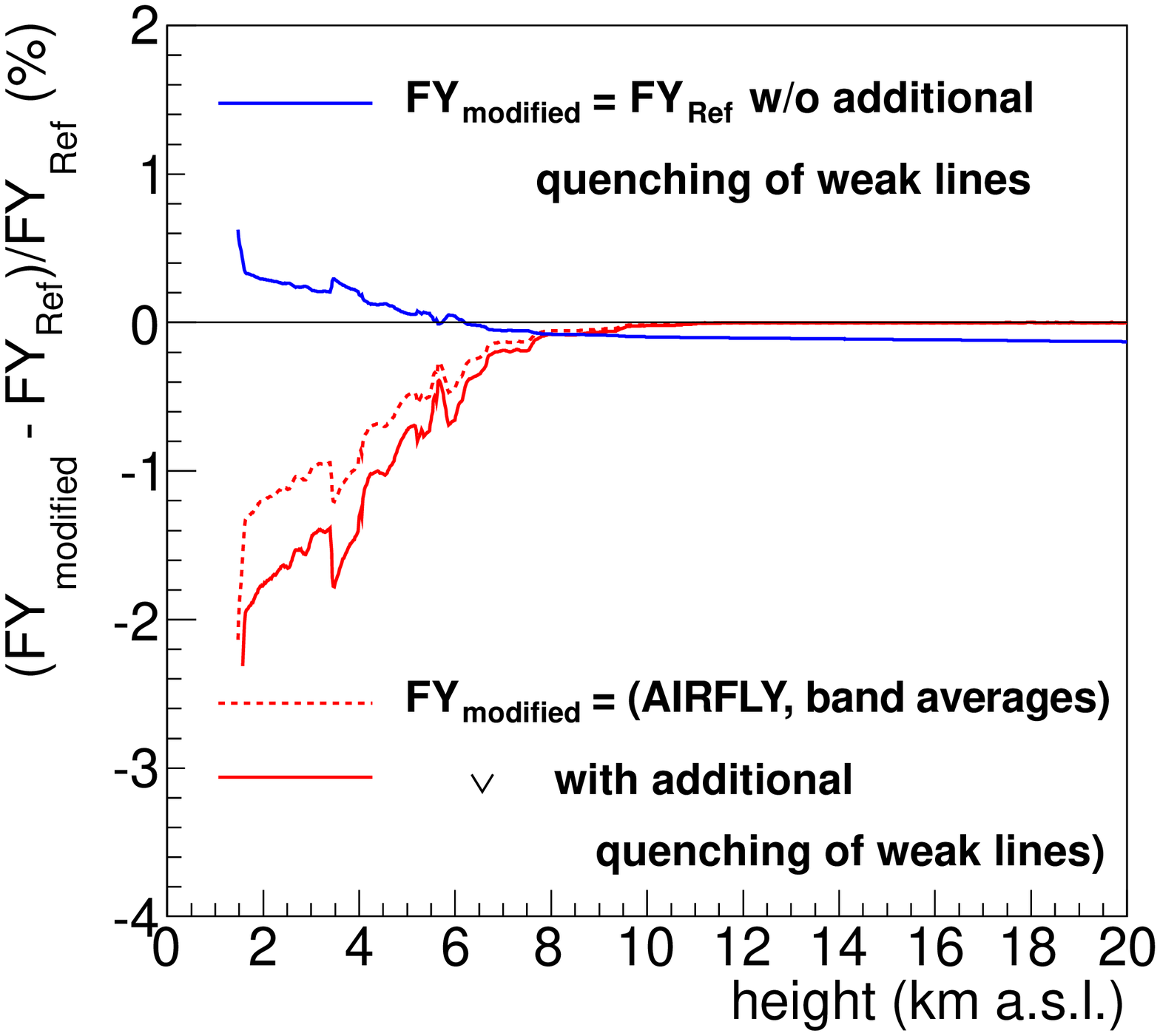}}
\end{minipage}
\hfill
\begin{minipage}[t]{0.48\textwidth}
\centering
\resizebox{1.\columnwidth}{!}{
\includegraphics{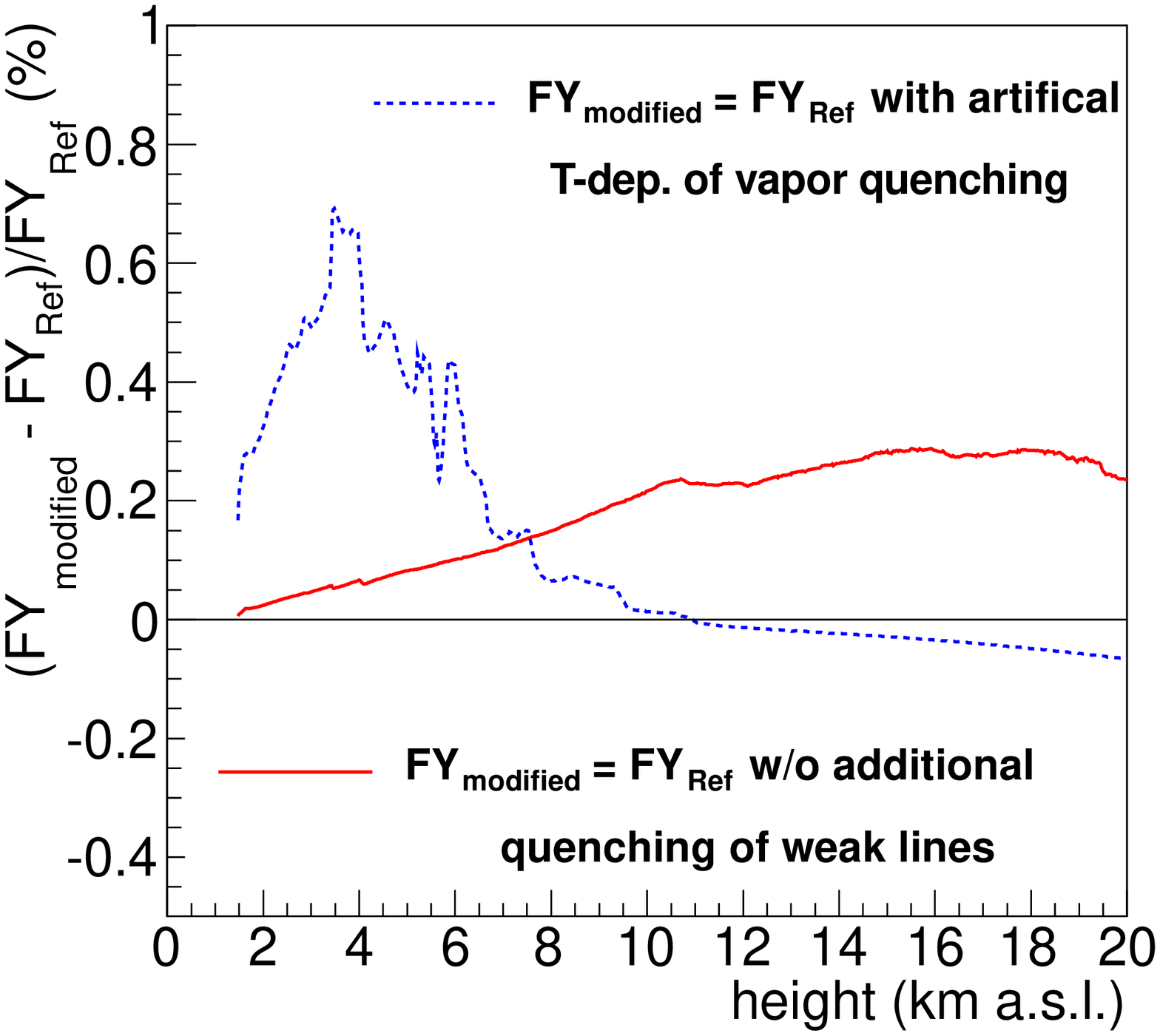}}
\end{minipage}
\caption{Left: Variations of \pprimevapor. Right: Variations of $\alpha$.}
\label{fig:variations}
\end{figure}

Some systematics about the choice of the suggested parameters for the 
reference description of the fluorescence emission are investigated and
displayed in Fig.~\ref{fig:variations}. For both, the quenching by 
water vapor and the temperature-dependent collisional cross sections,
the effect of applying estimated values to the weaker transitions are
presented.For the water vapor quenching, also another data set is 
shown for comparison (Fig.~\ref{fig:variations}, left, red curves).
In the case of temperature-dependent collisional cross sections, there
have been no measurements up to now to determine if there is also a
temperature-dependence for the de-excitation by water vapor molecules.
Thus, we applied an artificial dependence which is calculated with a 
factor of -5 to the known $\alpha$-values for water vapor to simulate
a quite extreme case. As can be seen in Fig.~\ref{fig:variations}, 
right, blue dotted curve, the overall variations are of minor importance.

\section{Application to Air Shower Reconstruction}
\label{sec:EASreco}
 
In the following, the reference description of the fluorescence emission
is applied to full air shower reconstructions. Here, we use an exemplary
set of high-quality hybrid events which have been observed during one year
by the Pierre Auger Observatory with both detector components, the fluorescence
detector and the surface detector for secondary
particles~\cite{AugerFD,AugerSD,AugerHybrid}. The reconstructions are
performed with the software framework \Offline~\cite{offline} taking into account all
necessary calibrations of the different detectors as well as actual atmospheric
conditions in terms of profiles of atmospheric state variables and
measured aerosol conditions~\cite{PesceICRC2011,GDASpaper,atmosPaper}.
The same set of air shower events is reconstructed in the same manner
several times, only the description of the fluorescence emission is
varied. To study how the fluorescence description affects the
reconstruction results, we focus here on the air shower observables
primary energy $E$ and position of shower maximum \xmax.

\subsection{Reconstruction results with the same absolute scaling}
\label{sec:reco_results}

\begin{figure}[tbp]
\hfill
\begin{minipage}[t]{0.48\textwidth}
\centering
\resizebox{1.\columnwidth}{!}{
\includegraphics{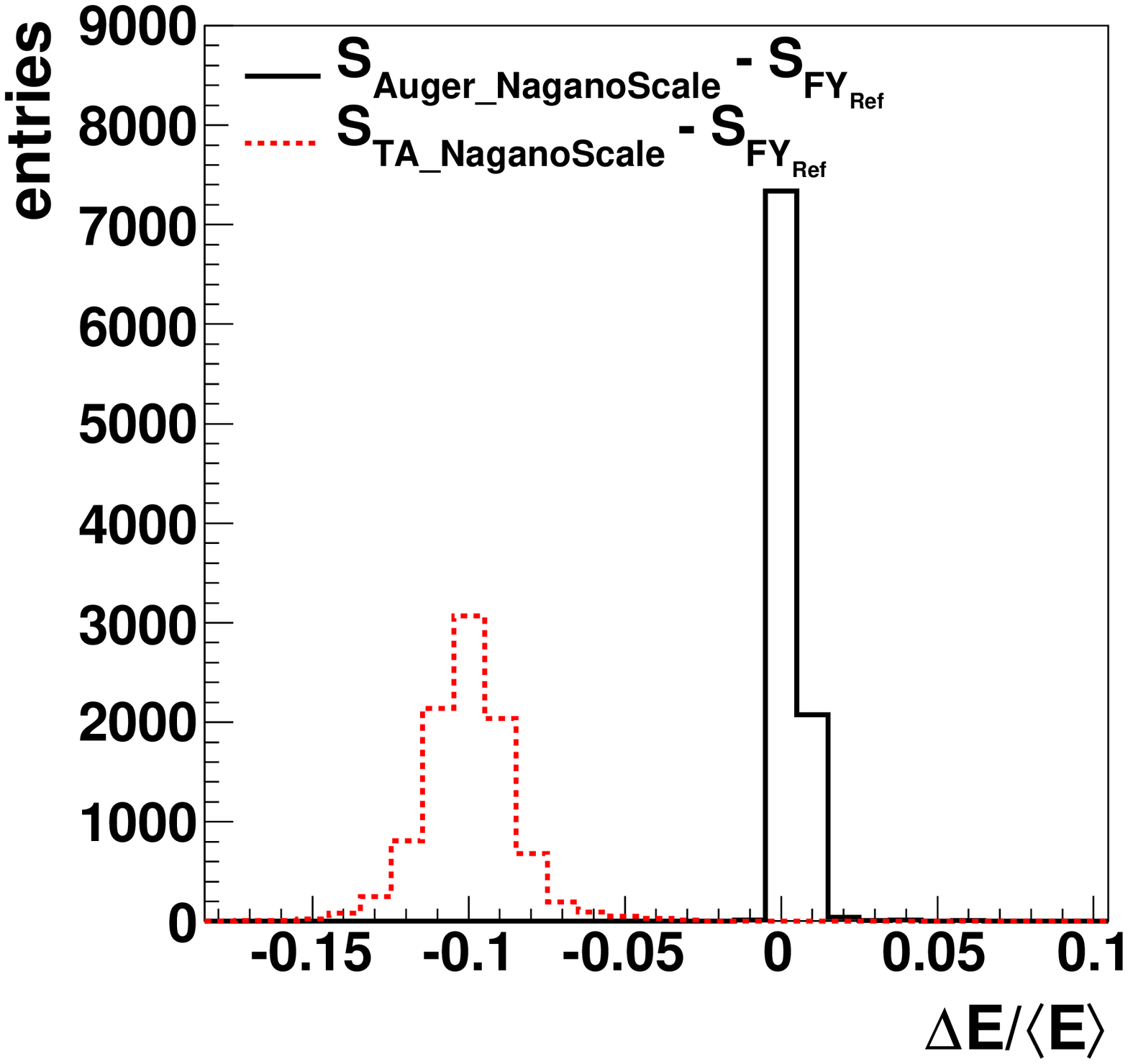}}
\end{minipage}
\hfill
\begin{minipage}[t]{0.48\textwidth}
\centering
\resizebox{1.\columnwidth}{!}{
\includegraphics{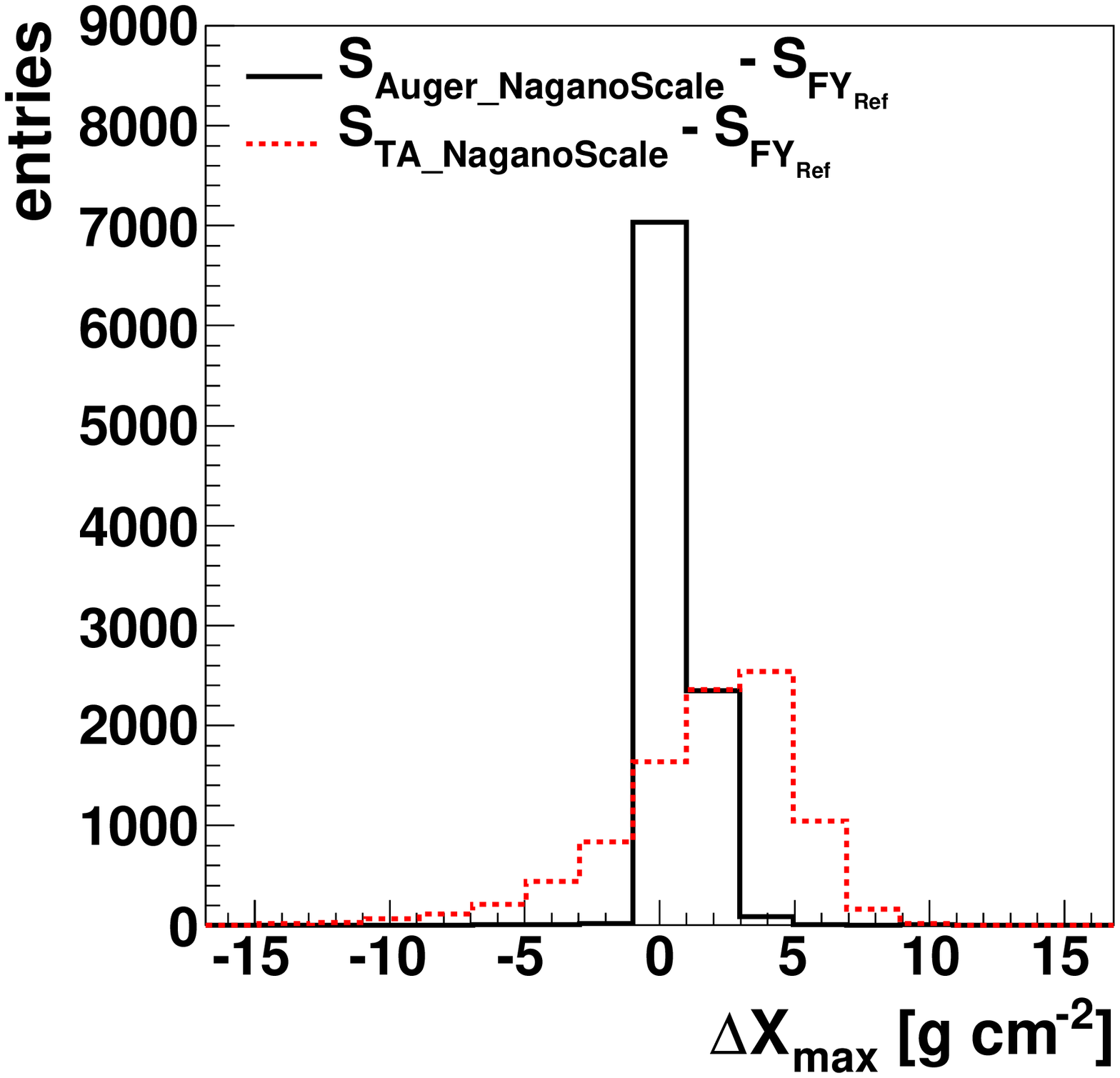}}
\end{minipage}
\caption{Left: Difference of reconstructed primary energy of air showers.
\setAugNag - \setRefyield: Mean 0.34\%, RMS 1.22\%; 
\setTANag - \setRefyield: Mean -9.98\%; RMS 1.76\%. \newline
Right: Difference of reconstructed position of shower maximum of air showers.
\setAugNag - \setRefyield: Mean 0.71~\gcmsq, RMS 1.25~\gcmsq; 
\setTANag - \setRefyield: Mean 1.63~\gcmsq; RMS 3.56~\gcmsq.}
\label{fig:deltas_sameabs}
\end{figure}

For this reconstruction study, we use in all calculations the same
absolute scaling of the fluorescence yield which is taken from Nagano
et al.~\cite{Nagano2004}. The fluorescence emission at 800~hPa and 
293~K of the 337.1~nm line is 6.38~$\gamma_{337}$/MeV. The reconstruction
set based on the reference fluorescence description \setRefyield
is taken as the 'standard' and uses the data given in Tab.~\ref{tab:refYield}.
The reconstruction set of the same data which uses the standard procedure of 
the Auger Observatory is labeled with \setAugNag. Furthermore, the
fluorescence description used in the TA experiment is implemented into
the reconstruction of these air shower events obtained by the 
Auger Observatory and also scaled by the same fluorescence yield 
of the 337.1~nm line, \setTANag. In Fig.~\ref{fig:deltas_sameabs}, the
different sets of reconstructions are compared to that with the 
reference fluorescence description. On the left, the difference of the 
reconstructed primary energy of the air showers $E$ is given and on 
the right, the difference of the reconstructed position of shower 
maximum \xmax is shown. It should be mentioned that in the case of the 
TA-reconstruction, no temperature-dependent collisional 
cross sections and no humidity quenching is taken into account.
The intensity spectrum is taken from the FLASH experiment, as 
shown in Fig.~\ref{fig:intensities}.

\begin{figure}[htbp]
\hfill
\begin{minipage}[t]{0.48\textwidth}
\centering
\resizebox{1.\columnwidth}{!}{
\includegraphics{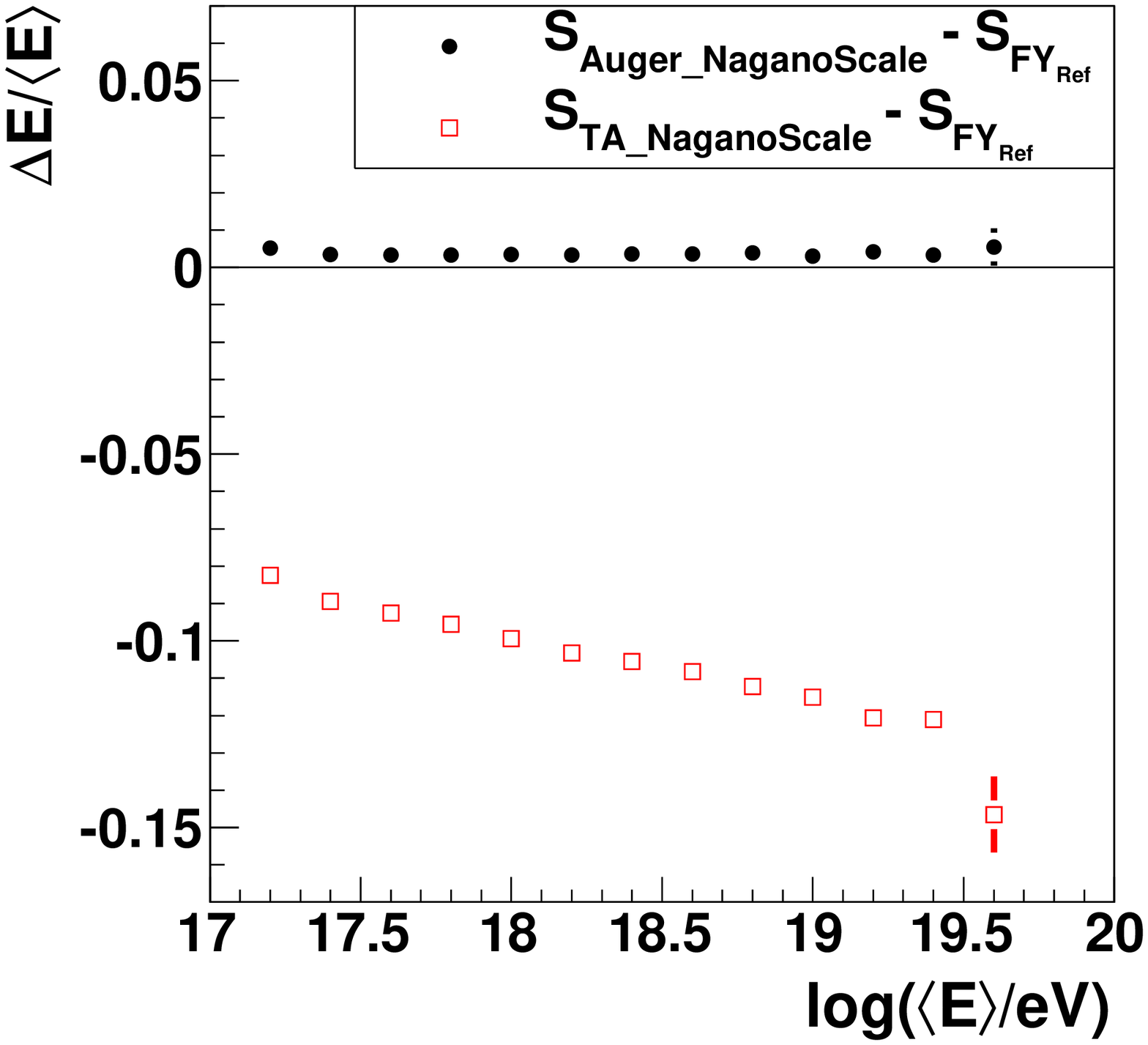}}
\end{minipage}
\hfill
\begin{minipage}[t]{0.48\textwidth}
\centering
\resizebox{1.\columnwidth}{!}{
\includegraphics{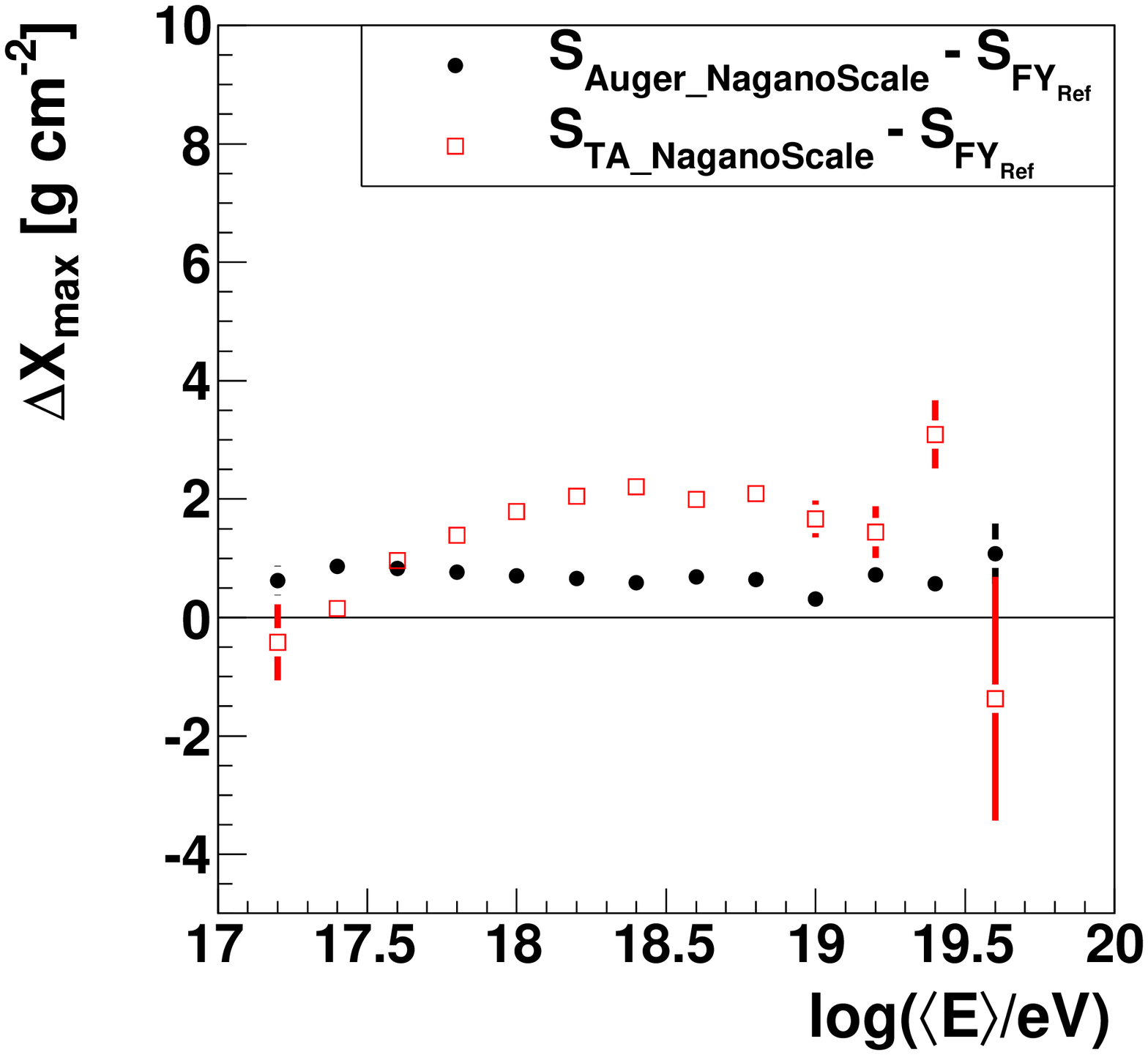}}.
\end{minipage}
\begin{minipage}[t]{0.48\textwidth}
\centering
\resizebox{1.\columnwidth}{!}{
\includegraphics{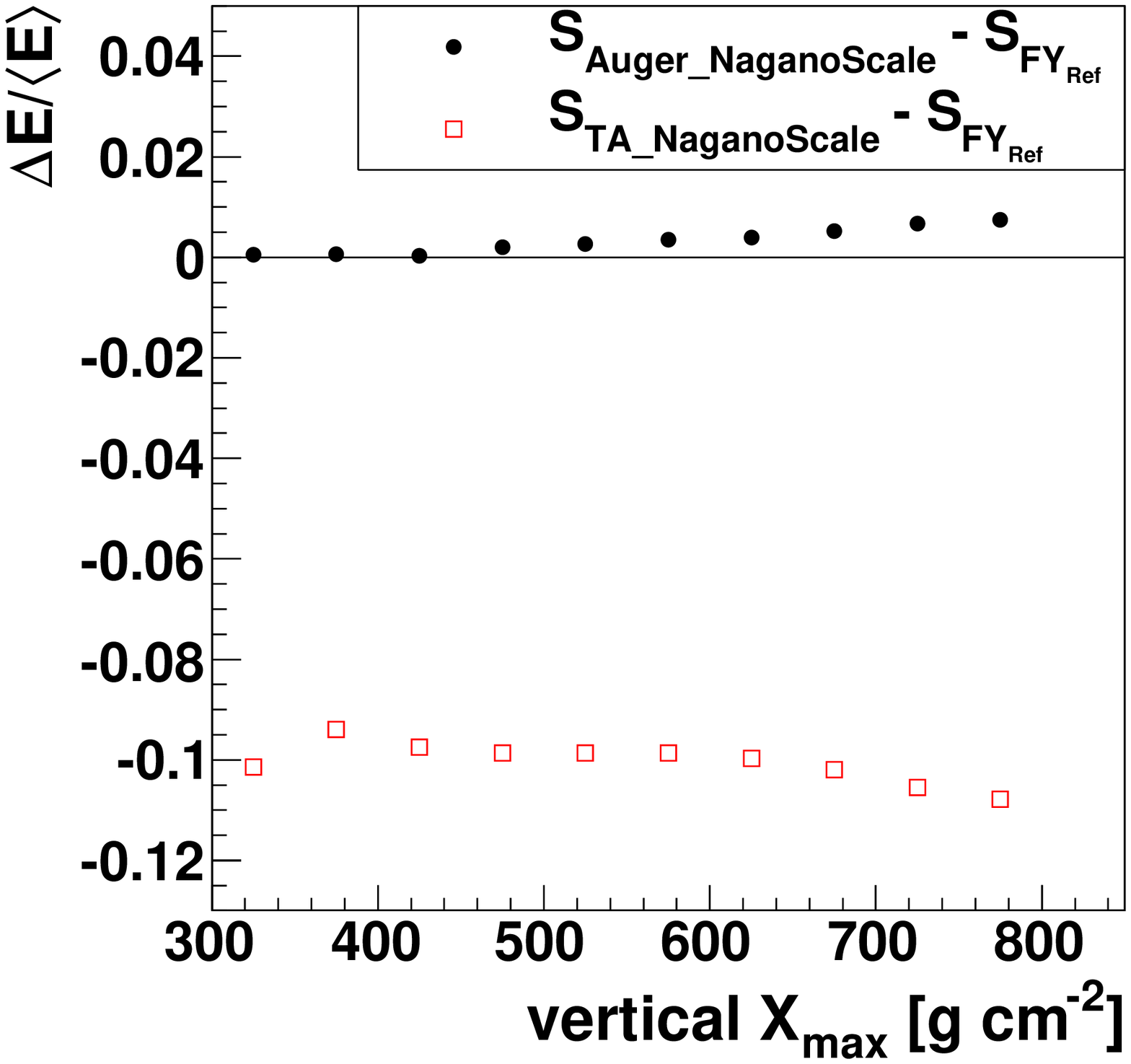}}
\end{minipage}
\hfill
\begin{minipage}[t]{0.48\textwidth}
\centering
\resizebox{1.\columnwidth}{!}{
\includegraphics{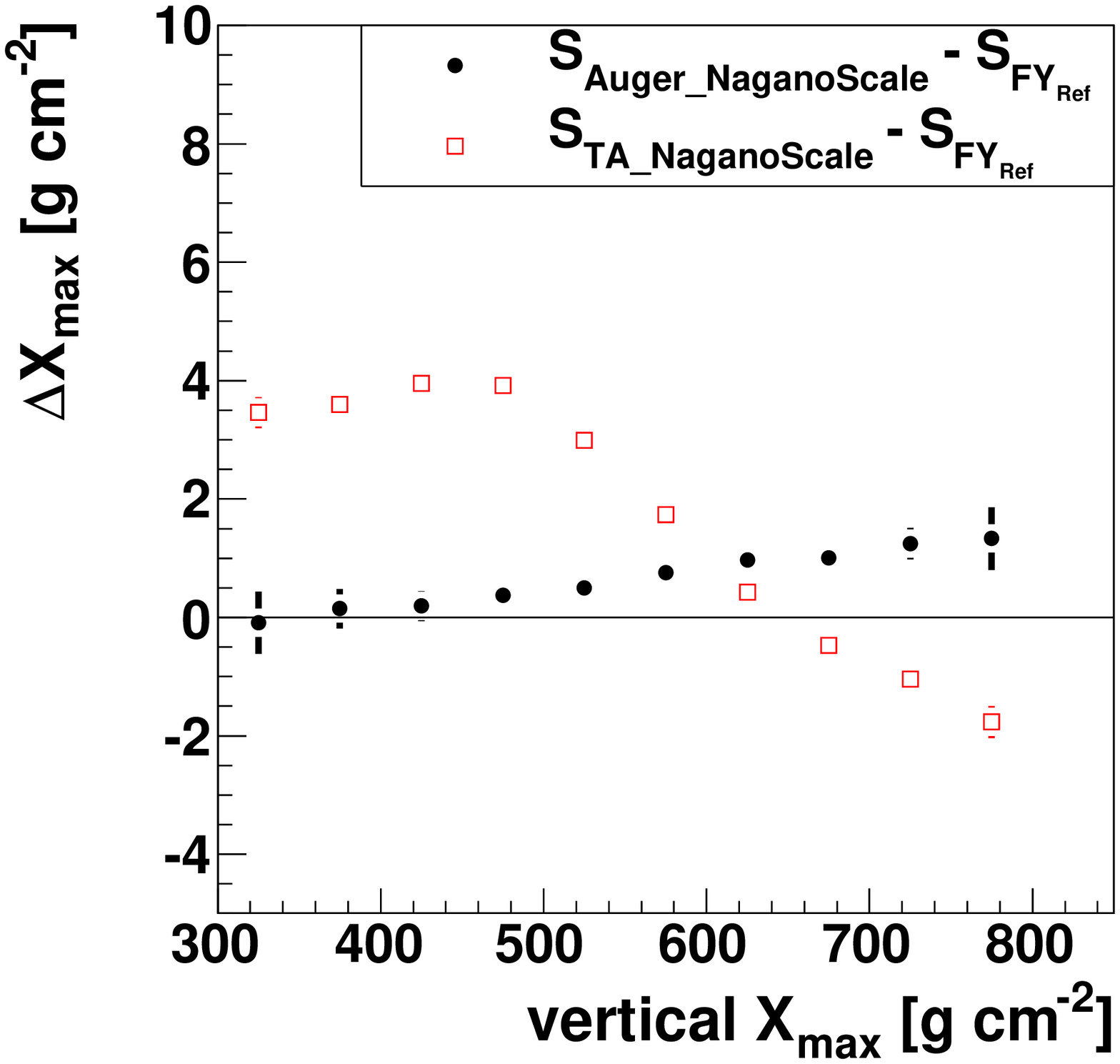}}
\end{minipage}
\begin{minipage}[t]{0.48\textwidth}
\centering
\resizebox{1.\columnwidth}{!}{
\includegraphics{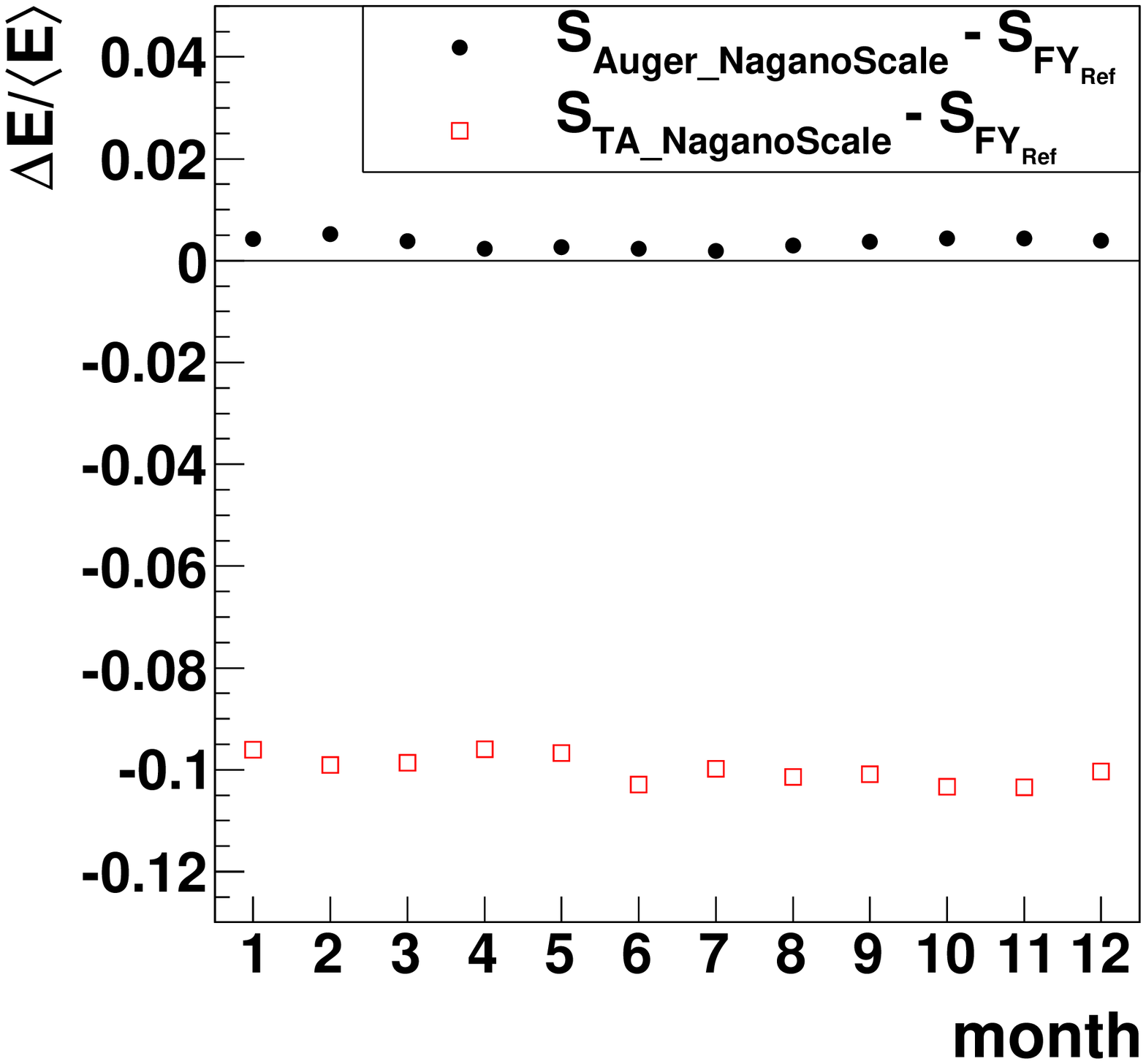}}
\end{minipage}
\hfill
\begin{minipage}[t]{0.48\textwidth}
\centering
\resizebox{1.\columnwidth}{!}{
\includegraphics{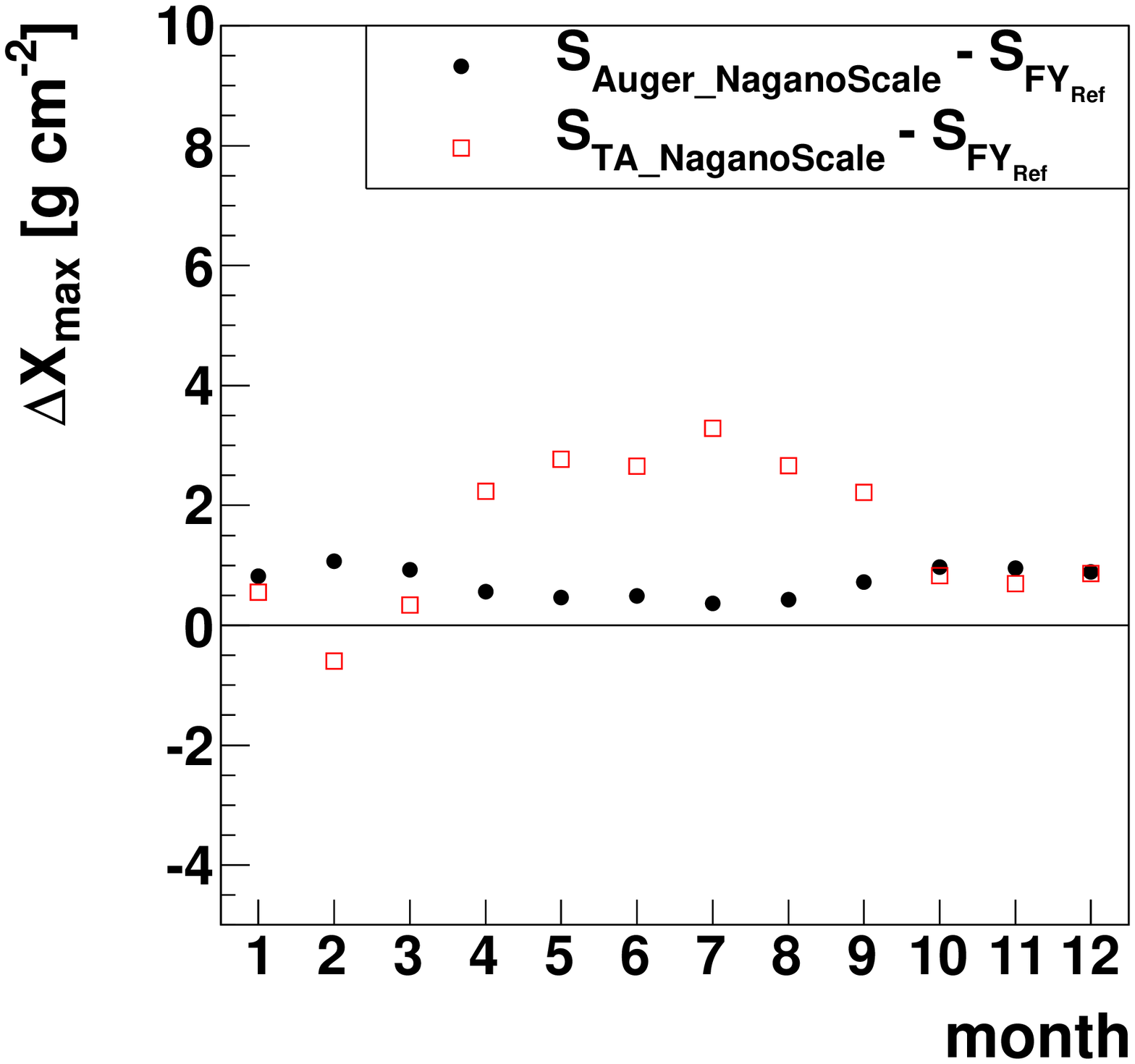}}
\end{minipage}
\caption{Left: Difference of reconstructed primary energy of air showers.
Right: Difference of reconstructed position of shower maximum of air showers.
From top to bottom: in dependence on primary energy, vs.\ vertical \xmax,
seasonal variations.}
\label{fig:EAS_systematics}
\end{figure}

In the following, some possible systematics are investigated. The 
fluorescence emission dominates the energy scale of the reconstructions,
so the differences of the reconstruction sets are plotted vs.\ $E$ in 
the upper part of Fig.~\ref{fig:EAS_systematics}. While the dependence
on $E$ is very flat and small for \setAugNag both for the energy and 
\xmax reconstruction, a considerable dependence is found for 
\setTANag in particular for the reconstructed energy. 

The atmospheric dependences are studied by analyzing the reconstructions
vs.\ vertical \xmax of the air shower 
(Fig.~\ref{fig:EAS_systematics}, middle row) and vs.\ month
(Fig.~\ref{fig:EAS_systematics}, bottom row). In the first case,
the influence of different atmospheric profiles of the state variables on
e.g.\ deeply or shallowly penetrating air showers are emphasized. There is
hardly any systematics for the reconstructed energy, but the
reconstruction of \xmax is affected. Because of the missing
temperature-dependent collisional cross sections, very shallow
air showers are reconstructed with a $\approx$ 4~\gcmsq larger
\xmax for \setTANag than for \setRefyield. The influence of 
the quenching by water vapor is emphasized for deeply penetrating
air showers and small modifications of the reconstructed \xmax are
found for both comparisons. The seasonal dependence is displayed
on a monthly basis. Also here, no significant effects are visible
for the reconstructed energy. For the reconstruction set \setTANag,
a noticeable modulation with seasons can be seen for the 
reconstructed position of shower maximum. 

\begin{figure}[tbp]
\hfill
\begin{minipage}[t]{0.48\textwidth}
\centering
\resizebox{1.\columnwidth}{!}{
\includegraphics{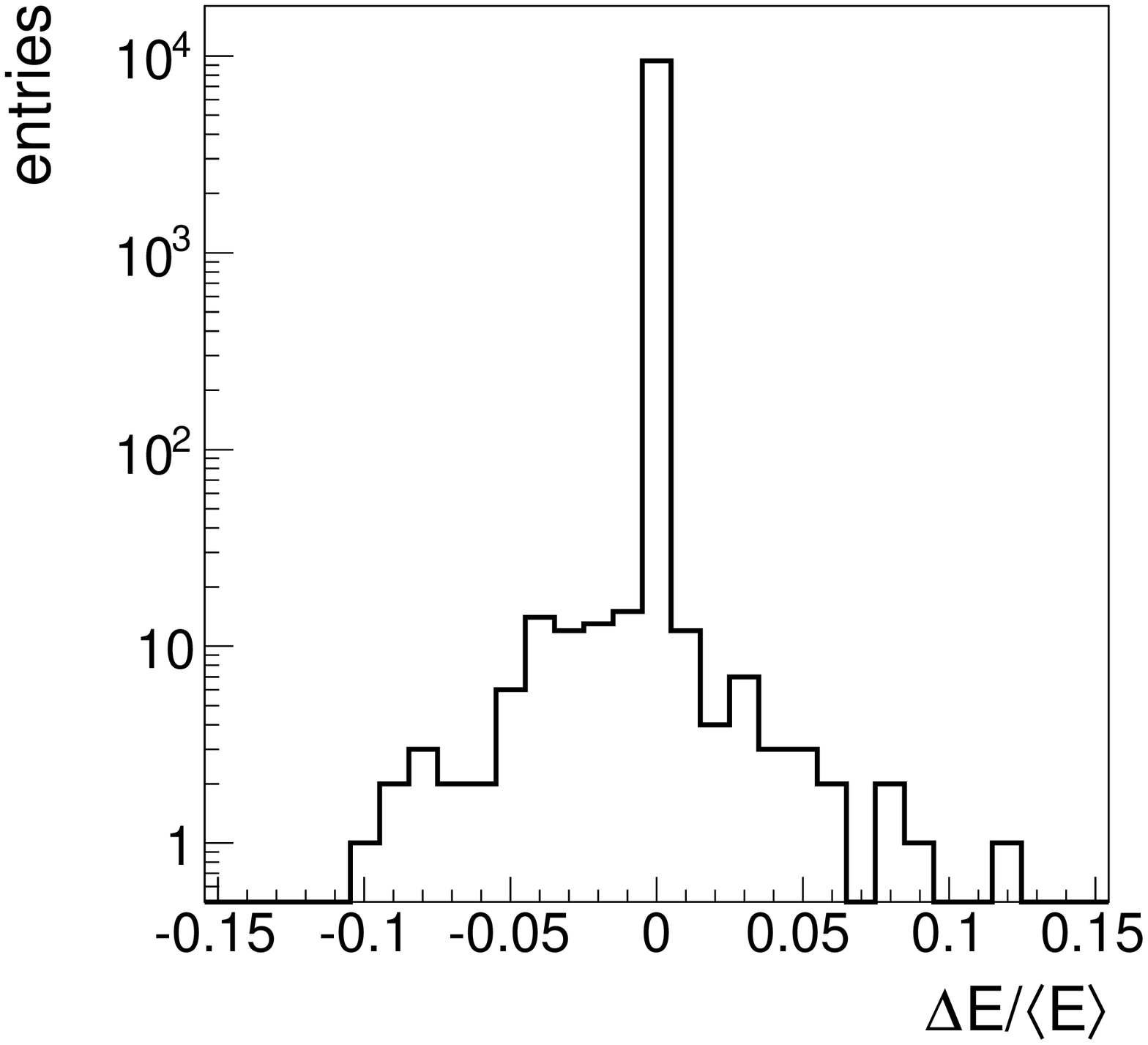}}
\end{minipage}
\hfill
\begin{minipage}[t]{0.48\textwidth}
\centering
\resizebox{1.\columnwidth}{!}{
\includegraphics{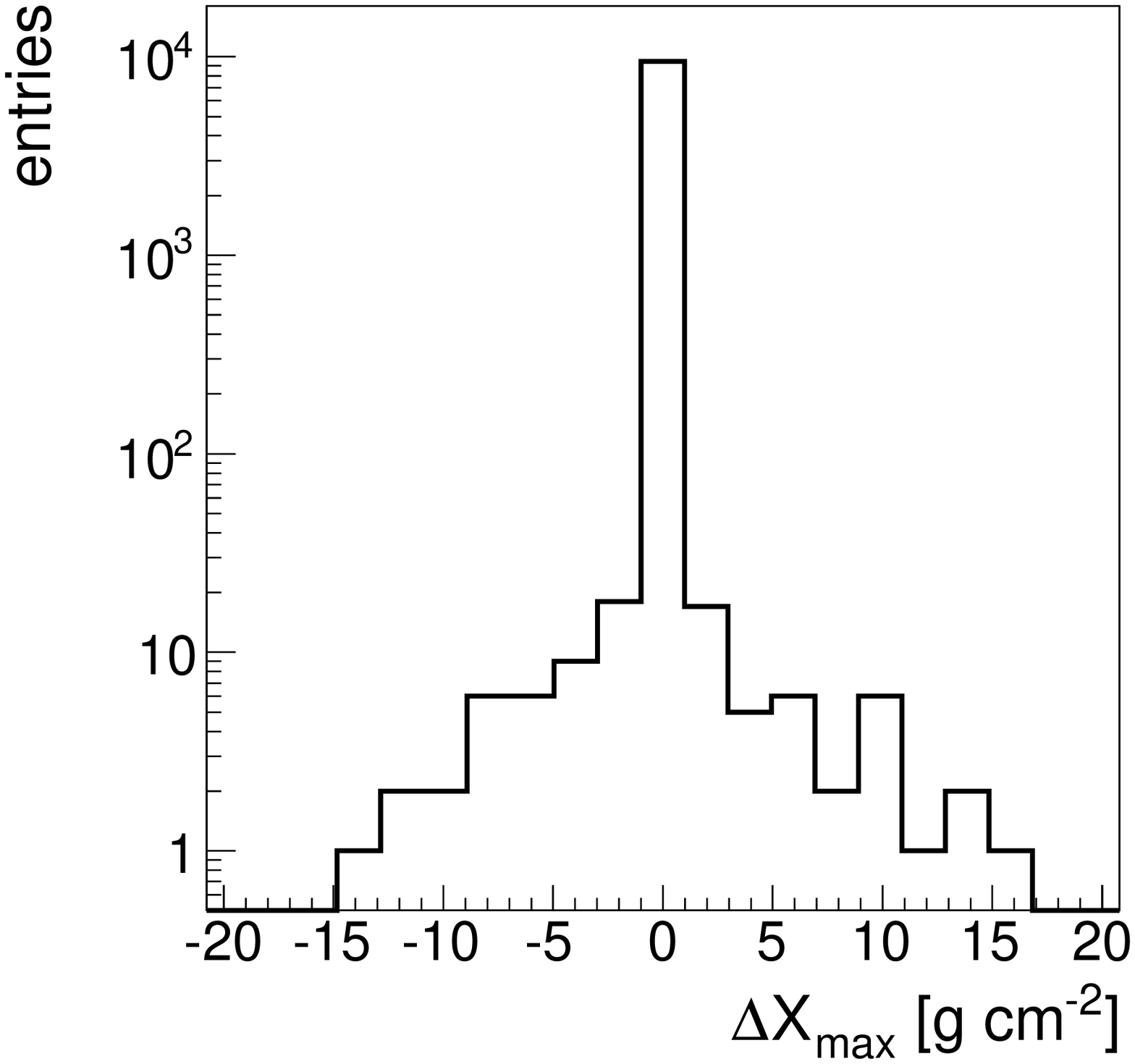}}
\end{minipage}
\caption{Difference of air shower reconstructions for the case
of ignoring the additional temperature and humidity dependences. 
Left: Primary energy.
Right: Position of shower maximum.}
\label{fig:deltas_variations}
\end{figure}

Even though some modifications of the fluorescence emission are
recognizable for the case of ignoring the additional temperature 
and humidity dependences (cf.\ Fig.~\ref{fig:variations}), the 
overall effect to full air shower reconstructions is negligible. 
Only very few events have differences in the reconstructed primary 
energy, left part of Fig.~\ref{fig:deltas_variations}, where the 
mean of the distribution is at -0.08\% with an RMS of 0.7\%. A
similar picture is given for the reconstructed \xmax, right part
of Fig.~\ref{fig:deltas_variations}, where the 
mean of the distribution is at -0.003~\gcmsq with an RMS of 
1.02~\gcmsq.

\subsection{Discussion of different absolute scalings}
\label{sec:abs_scaling}

In this part of the reconstruction study, different absolute 
scalings of the fluorescence yield are used. Typically, the absolute
yield is determined for the main emission line at 337.1~nm and the
spectral intensities of all other transitions are measured relatively
to that main emission. For the reference
description presented here, \setRefyield, the absolute yield from
Nagano et al.~\cite{Nagano2004} is applied, together with the 
spectral intensities of AIRFLY~\cite{airfly1}. Recently, the 
AIRFLY Collaboration has presented an own absolute yield for the 
337.1~nm line, which is 7.07~$\gamma_{337}$/MeV at 800~hPa and 293~K,
corresponding to 5.60(1)~$\gamma_{337}$/MeV at 1013~hPa and 
293~K~\cite{airfly6,bohacovaKA}. In the following, reconstructions
using this absolute yield together with the reference description 
are labeled with \setRefyieldAirfly. The TA experiment 
typically applies in its reconstructions an intensity spectrum provided 
by the FLASH experiment for the entire wavelengths range between
300 and 420~nm together with an absolute scale from Kakimoto et 
al.\ for the total emission between 300 and 400~nm~\cite{kakimoto}.
Converting this into an absolute yield of the main emission 
corresponds to 5.40~$\gamma_{337}$/MeV at 800~hPa and 293~K. The 
reconstruction set with these data is indicated by \setTAFlash.
Very recently, the group of Ulrich et al.\ at the 
\emph{Technische Universit\"at M\"unchen} - TUM also provided
an absolute yield of the 337.1~nm line~\cite{TUM4}. They derived
a value of 6.3~$\gamma_{337}$/MeV at 1000~hPa and 293~K, which
corresponds to 8.3~$\gamma_{337}$/MeV at 800~hPa and 293~K. The 
reconstruction set with this absolute yield, the spectral intensities
also from TUM (cf.\ Fig.~\ref{fig:intensities}), and all other
parameters as given for the reference yield is labeled with
\setRefyieldTUM. Finally, these values can be compared with a 
theoretical calculation performed by J.~Rosado and F.~Arqueros.
With a sophisticated Monte Carlo simulation, they derive a
value of 6.3~$\gamma_{337}$/MeV at 1013~hPa and 
293~K~\cite{rosado,rosadoUHECR}.

\begin{figure}[tbp]
\hfill
\begin{minipage}[t]{0.48\textwidth}
\centering
\resizebox{1.\columnwidth}{!}{
\includegraphics{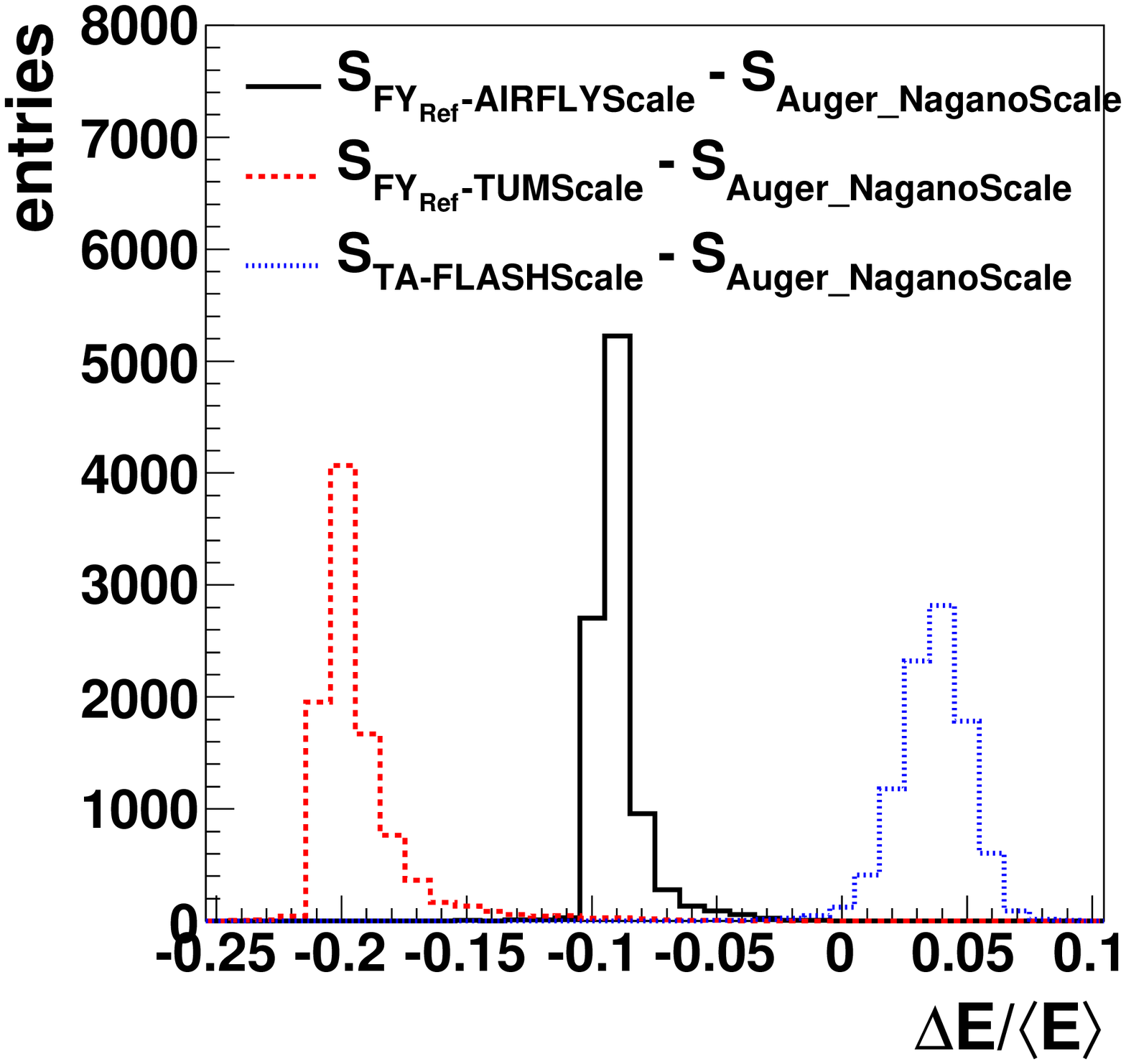}}
\end{minipage}
\hfill
\begin{minipage}[t]{0.48\textwidth}
\centering
\resizebox{1.\columnwidth}{!}{
\includegraphics{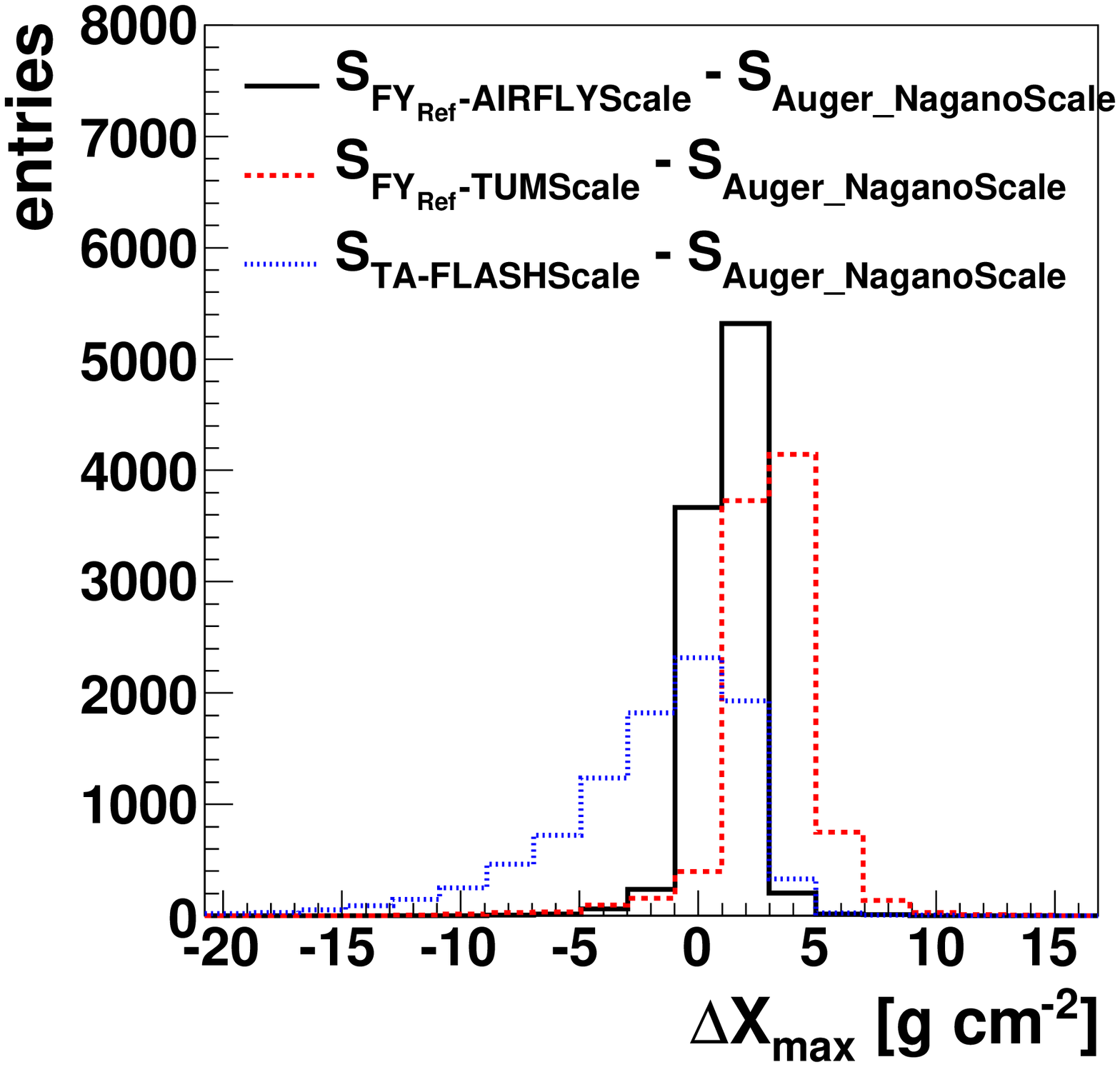}}
\end{minipage}
\caption{Left: Difference of reconstructed primary energy of air showers.
Right: Difference of reconstructed position of shower maximum of air showers.
In these figures, the different absolute yields are applied.}
\label{fig:deltas_diffabs}
\end{figure}

Applying the different absolute scales to the same air shower reconstruction
as presented in Sec.~\ref{sec:reco_results}, the differences for the 
reconstructed energy and \xmax can be analyzed, see Fig.~\ref{fig:deltas_diffabs}.
The mean energy of \setRefyieldAirfly is shifted by -9.0\% as compared with
the reconstruction of the Auger Observatory \setAugNag with an RMS of 1.5\% 
(solid black line of Fig.~\ref{fig:deltas_diffabs}, left). For \xmax, the 
mean is shifted by 1.1~\gcmsq (RMS 1.4~\gcmsq)
(solid black line of Fig.~\ref{fig:deltas_diffabs}, right).
Applying the scale derived from the TUM working group, the modification
is even more extreme (dashed red line in Fig.~\ref{fig:deltas_diffabs}).
The mean reconstructed energy is shifted by -19.3\% (RMS 2.4\%) and
the mean position of shower maximum by 3.0~\gcmsq (RMS 2.2~\gcmsq).
Comparing the reconstruction applying the fluorescence description
of the TA experiment with that of the Auger Observatory gives
a mean energy shift of 3.6\% (RMS 1.8\%). The mean \xmax is shifted
by -2.0~\gcmsq with a large tail represented by an RMS of 4.1~\gcmsq.

\section{Conclusions and Outlook}
\label{sec:concl}

We presented a first suggestion of a common reference description
of the fluorescence emission in air which is widely used for
high-energy cosmic ray observations. For the spectral and 
atmospheric dependences, detailed parameters could be derived 
from different experimental studies. Smaller variations of these
data are still possible because of some pending publication
of experimental data and some improvements in our algorithms.
However, the overall description of the altitude-dependent
fluorescence emission has reached an appropriate level for 
air shower reconstructions. The absolute scaling is still
to debate but we hope to revive the discussion with all parties
to find a reasonable decision soon.

\section*{Acknowledgments}

First of all, we want to thank the organizers of the inspiring 
UHECR2012 symposium. Furthermore, we thank all participants of
the 8$^{th}$ Air Fluorescence Workshop, who gave us the mandate
to develop a common reference description of the fluorescence 
emission. In particular, we acknowledge all colleagues of the 
'fluorescence community' who provided us their data and 
details about their experiments and helped us with fruitful 
discussion. And we are grateful to all (our) cosmic ray
collaborations applying the fluorescence technique but 
especially the Telescope Array Collaboration and the 
Pierre Auger Collaboration for allowing us to use real
event data and the reconstruction framework \Offline.
Part of this work was funded by the \emph{Bundesministerium
f\"ur Bildung und Forschung} (BMBF) and by the 
\emph{Deutsche Forschungsgemeinschaft} (DFG), Germany.

\end{document}